\documentclass[journal,10pt]{IEEEtran}
\usepackage{cite}
\ifCLASSINFOpdf
   \usepackage[pdftex]{graphicx}
\else
   \usepackage[dvips]{graphicx}
\fi
\usepackage[cmex10]{amsmath}
\usepackage{amsfonts,amssymb}
\usepackage{color}

\newtheorem{proposition}{Proposition}
\newtheorem{algorithm}{Algorithm}
\newcommand{\prox}[2]{\mathrm{prox}_{#1} \left(#2\right)}
\DeclareMathOperator*{\argmin}{arg\,min}

\renewcommand{\IEEEQED}{\IEEEQEDopen}

\begin{document}
\title{Symbol Detection for Frame-Based\\ Faster-than-Nyquist Signaling\\via Sum-of-Absolute-Values Optimization}

\author{Hampei~Sasahara,~\IEEEmembership{Student Member,~IEEE,}
        Kazunori~Hayashi,~\IEEEmembership{Member,~IEEE,}
        and~Masaaki~Nagahara,~\IEEEmembership{Senior~Member,~IEEE}% <-this % stops a space
\thanks{The authors are with the Graduate School of Informatics, Kyoto University,
	Kyoto, 606-8501~Japan~e-mail: (sasahara.h@acs.i.kyoto-u.ac.jp;kazunori@i.kyoto-u.ac.jp;nagahara@i.kyoto-u.ac.jp).}% <-this % stops a space
\thanks{Manuscript received April 19, 2005; revised September 17, 2014.}}


\markboth{IEEE Communications Letters,~Vol.~X, No.~Y, Month~2015}%
{Shell \MakeLowercase{\textit{et al.}}: Bare Demo of IEEEtran.cls for Journals}



\maketitle

\begin{abstract}
In this letter, we propose a new symbol detection method for faster-than-Nyquist signaling (FTNS) systems.
Based on frame theory, we formulate a symbol detection problem as a under-determined linear equation on a finite set.
The problem is reformulated as a sum-of-absolute-values (SOAV) optimization that can be efficiently solved by the fast iterative shrinkage thresholding algorithm (FISTA).
The proximity operator for the convex optimization is derived analytically.
Simulation results are given to show that the proposed method can successfully detect symbols in faster-than-Nyquist signaling systems and has lower complexity in terms of computation time.
\end{abstract}


% \fi
\IEEEpeerreviewmaketitle



\section{Introduction}
\IEEEPARstart{F}{aster}-than-Nyquist signaling (FTNS), which was proposed by Mazo in 1975, is a framework to transmit signals exceeding the Nyquist rate~\cite{Mazo1975}.
It is shown that a 24.7 \% faster symbol rate than the Nyquist rate can be achieved without performance loss in terms of the minimum Euclidian distance for binary symbols and sinc pulses.
Moreover, the capacity of FTNS is higher than conventional Nyquist rate signaling from an information theoretic viewpoint~\cite{Fredrik2009}.
For these reasons, FTNS has been drawing great attention as a new method realizing high-speed data transmission~\cite{FTN}.
%Many researchers have proposed and studied systems based on the FTNS framework,
%for instance, multicarrier systems~\cite{Rusek2005}.

Nyquist discovered that if the symbol period is shorter than the inverse of the pulse bandwidth,
then inter-symbol interference (ISI) is unavoidable~\cite{Nyquist1928}.
The key idea of actualization of FTNS is that, by appropriate signal processing, one can sufficiently reduce ISI caused by a shorter symbol period than that of the Nyquist criterion requires.
When ISI happens due to the fast symbol rate, the sifted pulses become linearly dependent and it is impossible to recover arbitrary symbols in the real axis or the complex plane.
However, as pointed out in~\cite{Fang2009}, considering the fact that candidates of transmitted symbols are elements of a finite set in digital communication,
we might perfectly reconstruct symbols even when the symbol rate is higher than the Nyquist rate.
This desirable situation can be attained when transmission pulses constitute Weyl-Heisenberg frames and
consequently, we can realize no loss FTNS using frame theory~\cite{Fang2009}.
Furthermore, it is shown that a system with frames is more robust against doubly selective fading than a system with conventional orthogonal pulses~\cite{Kozek1998,Matz2007}.
Because of these merits, we investigate the frame-based FTNS \cite{Strohmer01} in this letter.

%As shown by H. Nyquist in~\cite{Nyquist1928}, $TW \geq 1$ is a necessary condition for no inter-symbol interference (ISI),
%where $T$ [sec] and $W$ [Hz] are a symbol period and the frequency bandwidth of a modulation pulse respectively.
%This implies that pulses constitute Riesz Basis, i.e., they are linearly independent, from the viewpoint of functional analysis.
%On the other hand, when $TW < 1$, pulses become linearly dependent and it is impossible to recover arbitrary symbols in the real axis or the complex plane.
%However, as pointed out in~\cite{Fang2009}, considering the fact that candidates of transmitted symbols are elements of a finite set in digital communication,
%we might perfectly reconstruct symbols even when the symbol rate is higher than the Nyquist rate.
%This desirable situation can be attained when transmission pulses constitute Weyl-Heisenberg (W-H) frames and
%consequently, we can realize no loss FTNS using frame theory~\cite{Fang2009}.
%Furthermore, it is shown that a system with W-H frames is more robust against doubly selective fading than a system with conventional orthogonal pulses~\cite{Kozek1998,Matz2007}.
%Taking these merits, we consider frame-based FTNS \cite{Strohmer01} in this letter.

In practice, symbol detection is one of the fundamental issues to realize FTNS.
As standard detection methods for conventional FTNS, maximum likelihood (ML) approaches, such as the Viterbi algorithm (VA) and the Bahl-Cocke-Jelinek-Raviv (BCJR) algorithm, have been proposed for uncoded FTNS~\cite{Anderson2009}.
For coded FTNS, turbo receivers have been proposed in~\cite{Prlja2008}.
One the other hand, as a detection scheme for the frame-based FTNS, the $\ell^{\infty}$ minimization method has been proposed and it is shown that the computational complexity of the method is lower than that of the above algorithms~\cite{Fang2014}.
However, further reduction of the complexity will be required for the application to practical systems.
%However, they are insufficient in terms of detection performance or computational burden.

In this letter, we propose a symbol detector for frame-based FTNS using \emph{sum of absolute values (SOAV)} optimization.
SOAV has been recently proposed as a discrete valued signal reconstruction method and is known to be very effective when the number of candidate discrete values is not so large~\cite{DSR}.
While only noiseless cases are considered in~\cite{DSR}, we extend the problem formulation into noisy cases.
The fast iterative shrinkage thresholding algorithm (FISTA)~\cite{Beck2009} is utilized as a solver for the optimization problem, since it requires much lower computational complexity.
To derive an algorithm based on FISTA, we give the closed form of a proximity operator for the SOAV optimization.

The remainder of this letter is as follows:
Section~\ref{Smsd} gives the system model considered in this letter and Section~\ref{Det} proposes a symbol detection scheme.
In Section~\ref{OA} we introduce FISTA and derive an associated proximity operator.
Simulation results are shown in Section~\ref{SR} to illustrate the effectiveness of the proposed method.
Section~\ref{conc} draws conclusions.

\section{System model}
\label{Smsd}
In this section, we explain the frame-based FTNS system considered in this letter.

Let us consider the following linear modulation:
\[
 x(t) = \sum_{n=1}^N x_n h_n(t),
\]
where $N \in \mathbb{N}$ is the number of modulation waveforms,
$\{x_n\}_{n=1}^N \in \{+1,-1\}^N$ are independent identically distributed (i.i.d.) binary symbols,
and $\{h_n\}_{n=1}^N$ are the modulation pulses whose frequency bandwidth is limited to $W$.
The signal through an additive white Gaussian noise (AWGN) channel is
\[
 \begin{array}{ccl}
  y(t) & = & x(t) + w(t) \\
  & = & \displaystyle{\sum_{n=1}^N x_n h_n(t) + w(t),} \\
 \end{array}
\]
where $w(t)$ is the AWGN with zero mean and power spectral density $N_0$.
We assume that the symbol period is set to be $T$.
Let $M$ be the dimension of the time-frequency space occupied by $x(t)$.
$\{\phi_i \}_{i=1}^M$ is set to be an orthonormal basis for such a time-frequency space.
Define $y_m \triangleq \langle y(t), \phi_m(t) \rangle, h_{m,n} \triangleq \langle h_n(t), \phi_m(t) \rangle$, and $w_m \triangleq \langle w(t),\phi_m(t) \rangle$ for $m=1,\ldots,M$ and $n=1,\ldots,N$,
where $\langle \cdot, \cdot \rangle$ denotes an inner product.
Then we obtain the following linear equation~\cite{Fang2014}:
\begin{equation}
 y = Hx + w,
 \label{under_determined}
\end{equation}
where $x \triangleq [x_1,\ldots,x_N]^{\top} ,y \triangleq [y_1,\ldots,y_M]^{\top}, H \triangleq (h_{m,n}),$ $w \triangleq [w_1,\ldots,w_M]^{\top}$, and $[\cdot]^{\top}$ represents the transpose.
$w$ is a zero mean Gaussian random vector with covariance $E\{w w^{\top}\} = (N_0/2)I_M$, where $E\{\cdot \}$ and $I_M$ denote the expectation and the $M$ dimension identity matrix respectively.
It is assumed that $h_n$ is generated by i.i.d variables with zero mean and variance $1/M$.
Then, the modulation matrix $H$ is modeled as a random matrix~\cite{Fang2014-2}.

Note that, for a conventional FTNS system using an ordinary matched-filter,
$H$ will be a Toeplitz convolution matrix,
while discrete valued signal reconstruction scheme with SOAV does not work well for such a structured matrix.
Accordingly, we can also find one of merits of frame-based FTNS system.
Note also that, we can employ not only binary phase shift keying (BPSK) but also quadrature phase shift keying (QPSK) for the baseband modulation.
When QPSK is used, let $\tilde{x}$ be modulated symbols, $\tilde{w}$ be complex-valued noise and $\tilde{H}$ be a complex-valued modulation matrix.
Define $x \triangleq [\mathrm{Re} \{\tilde{x}^{\top}\}, \mathrm{Im} \{\tilde{x}^{\top}\}]^{\top}$, $w \triangleq [\mathrm{Re} \{\tilde{w}^{\top}\}, \mathrm{Im} \{\tilde{w}^{\top}\}]^{\top}$,
and 
\[
 H \triangleq \left[
 \begin{array}{cc}
 \mathrm{Re}\{H\} & -\mathrm{Im} \{H\} \\
 \mathrm{Im} \{H\} & \mathrm{Re} \{H\} \\
 \end{array}\right],
\]
where $\mathrm{Re}\{\cdot\}$ and $\mathrm{Im}\{\cdot\}$ denote the real part and the imaginary part respectively.
Then we obtain the above system (\ref{under_determined}).

\section{Symbol detection via SOAV optimization}
\label{Det}

When the system employs faster-than-Nyquist signaling, i.e., $TW > 1$, then $M<N$.
%Thus, (\ref{under_determined}) becomes an under-determined system.
In a noiseless case, the original signal $x$ can be found from the constrained under-determined system:
\[
 Hz = y,\ \mathrm{s.t.}\ z \in \{ +1,-1\}^N,
\]
since the number of the symbol set is finite~\cite{Fang2014}.
Similarly, in a noisy case the problem can be reformulated as
\begin{equation}
 \min_{z \in \{+1,-1\}^N} \|y-Hz\|_2,
\label{prob_noisy}
\end{equation}
where $\|\cdot\|_2$ represents the $\ell^2$ norm of the vector.
These problems, however, have a combinatorial nature and the computation time becomes exponential.

To reasonably obtain the solution of (\ref{prob_noisy}), the $\ell^{\infty}$-minimization method has been proposed for binary symbol recovery~\cite{Fang2014}.
In a noisy case, this method considers the relaxed problem
\[
 \min_{z\in \mathbb{R}^N}\|z\|_{\infty} \ \mathrm{s.t.}\ \|y-Hz\|_2^2 \leq \varepsilon^2,
\]
where $\|\cdot\|_{\infty}$ is defined as the $\ell^{\infty}$ norm of the vector.
It can be solved by repeating the Newton's method and the solution approximates the solution of (\ref{prob_noisy}).
The computational complexity is less than Viterbi algorithm; nevertheless it is insufficient for practical needs.

To tackle with this difficulty, we consider to estimate $x$ by solving the following optimization problem:
\begin{equation}
 \begin{array}{l}
 \displaystyle{\min_{z \in \mathbb{R}^N} \frac{1}{2}\|z-1_N\|_1 + \frac{1}{2} \|z+1_N\|_1} \\
 \mathrm{s.t.}\ \|y - Hz\|_2^2 \leq \varepsilon^2, \\
 \end{array}
\label{soav}
\end{equation}
where $\|\cdot\|_1$ is the $\ell^1$ norm of the vector, $\varepsilon \in \mathbb{R},$ and $1_{N}$ is the $N$ dimension vector whose all components are $1$.
This problem formulation can be considered as a noisy version of the SOAV optimization considered in the reference~\cite{DSR}.
The problem is a convex optimization problem and can be efficiently solved by several algorithms.
In particular, we propose a proximal algorithm that has low computational complexity in the next section.

We can interpret (\ref{soav}) as follows:
Because the original signal $x$ has elements of $\{+1,-1\}$,
it is expected that about half elements of $x-1_{N}$ and $x+1_{N}$ are all zero, if $+1$ and $-1$ appear with equal probability in $x$.
Consequently, based on the idea of compressed sensing, we anticipate that the original signal can be obtained by finding a vector $z$ which satisfies the constraint and makes the $\ell^1$ norms of $z-1_{N}$ and $z+1_{N}$ small.

\section{Optimization algorithm}
\label{OA}

It is well known that the proximal algorithms, which are for solving convex optimization problems, have low computational complexity~\cite{prox-alg}.
In order to apply a proximal algorithm to the optimization problem of (\ref{soav}),
we give the closed form of a proximity operator for the problem in this section.

First, we reformulate (\ref{soav}) as the unconstrained optimization problem
\begin{equation}
 \min_{z \in \mathbb{R}^N} \lambda \|y-Hz\|_2^2 + \frac{1}{2}\|z-1_N\|_1 + \frac{1}{2} \|z+1_N\|_1,
 \label{soav2}
\end{equation}
where $\lambda$ is a positive number.
Note that for any $\epsilon$ there exists $\lambda$ such that the solution of (\ref{soav2}) becomes equal to the solution of (\ref{soav}).
Letting $f(z) \triangleq \lambda \|y-Hz\|_2^2$ and $g(z) \triangleq \frac{1}{2}\|z-1_N\|_1 + \frac{1}{2} \|z+1_N\|_1$,
we rewrite the problem as
\[
 \min_{z \in \mathbb{R}^N} f(z) + g(z).
\]
Here $f$ is convex and differentiable.
Thus if we have the proximity operator of $g$, then FISTA can be applied to the problem~\cite{Beck2009,Fixed-point}.

\begin{figure}[t]
\centering
\includegraphics[width=0.8\linewidth]{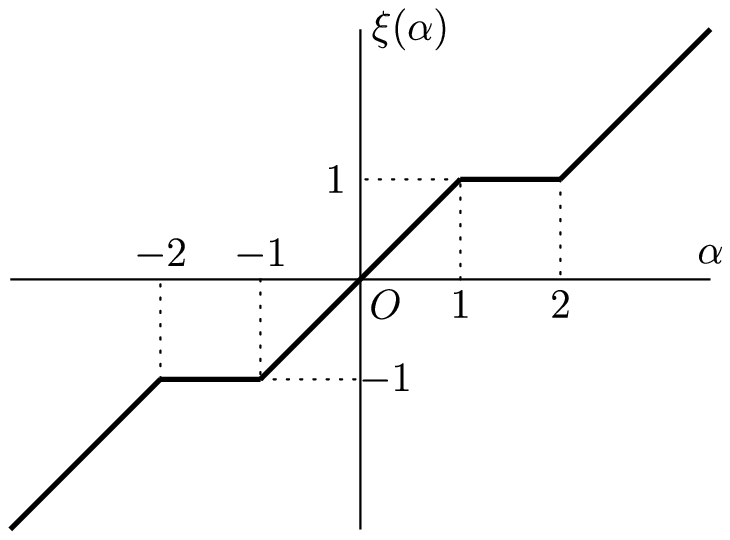}
\caption{Function $\xi(\alpha)$ for $\mathrm{prox}_g(z)$.}
\label{fig:prox}
\end{figure}

Define the proximity operator of $g$ as $\prox{g}{z} \triangleq \argmin_{u \in \mathbb{R}^N} \left\{ g(u) + \frac{1}{2} \|z-u\|_2^2 \right\}$.
Then we have the following proposition:
\begin{proposition}
Let $\xi: \mathbb{R} \to \mathbb{R}$ be
\[
 \xi(\alpha) \triangleq \left\{
 \begin{array}{ll}
 \alpha + 1 & \mathrm{if}\mbox{ }\alpha < {-2}, \\
 -1 & \mathrm{if}\mbox{ }{-2} \leq \alpha < {-1}, \\
 \alpha & \mathrm{if}\mbox{ }{-1} \leq \alpha < 1, \\
 1 & \mathrm{if}\mbox{ }1 \leq \alpha < 2, \\
 \alpha - 1 & \mathrm{if}\mbox{ }2 < \alpha \\
 \end{array}
 \right.
\]
(see also Fig.~\ref{fig:prox}).
Then we have
\[
 \prox{g}{z} = [\xi(z_1),\xi(z_2),\ldots,\xi(z_N)]^{\top},
\]
where $z_i$ is the $i$-th element of $z$.
\end{proposition}

\begin{IEEEproof}
The function $g$ can be written as
\[
 g(z) = \frac{1}{2}\sum_{i=1}^N(|z_i-1|+|z_i+1|),
\]
where $z_i$ is the $i$-th element of $z$.
From this, we have
\[
 \begin{array}{c}
 \displaystyle{\prox{g}{z} = \argmin_{u \in \mathbb{R}^N} \left\{\sum_{i=1}^N R(u_i,z_i)\right\},} \\
 R(\alpha,\beta) \triangleq |\alpha-1|+|\alpha+1|+|\beta-\alpha|^2, \\
 \end{array}
\]
where $\alpha,\beta \in \mathbb{R}$.
It is clear that $\mathrm{prox}_g(z)$ is obtained by minimizing each $R(u_i,z_i)$ independently.
Thus, we consider minimization of $R(\alpha,\beta)$ with $\alpha \in \mathbb{R}$ for fixed $\beta \in \mathbb{R}$.

If $\alpha \leq -1$, then we have
\[
 R(\alpha,\beta) = \{\alpha - (\beta+1)\}^2 + 2\beta - 1 =: R_1(\alpha,\beta).
\]
If $-1 \leq \alpha \leq 1$, then we have
\[
 R(\alpha,\beta) = (\alpha-\beta)^2 + 2 =: R_2(\alpha,\beta).
\]
If $\alpha \geq 1$, then we have
\[
 R(\alpha,\beta) = \{\alpha - (\beta-1)\}^2 - 2\beta - 1 =: R_3(\alpha,\beta).
\]
In summary, we have
\[
 R(\alpha,\beta) = \left\{
 \begin{array}{ll}
 R_1(\alpha,\beta), & \mathrm{if}\mbox{ }\alpha \in (-\infty,-1], \\
 R_2(\alpha,\beta), & \mathrm{if}\mbox{ }\alpha \in [-1,1], \\
 R_3(\alpha,\beta), & \mathrm{if}\mbox{ }\alpha \in [1,\infty). \\
 \end{array}\right.
\]

Based on this, we calculate the proximity operator.
We consider the cases:
\begin{enumerate}
\item $\beta < -2$,
\item $-2 \leq \beta < -1$,
\item $-1 \leq \beta < 1$,
\item $1 \leq \beta < 2$,
\item $2 \leq \beta$.
\end{enumerate}
When $\beta < -2$, we have
\[
 \begin{array}{c}
 \argmin_{\alpha \in (-\infty,-1]} \{R_1(\alpha,\beta)\} = \beta + 1, \\
 \argmin_{\alpha \in [0,1]} \{R_2(\alpha,\beta)\} = -1, \\
 \argmin_{\alpha \in [1,\infty)} \{R_3(\alpha,\beta)\} = 1. \\
 \end{array}
\]
Since $R_1(\beta+1,\beta) \leq R_2(-1,\beta) \leq R_3(1,\beta)$, we have
\[
 \argmin_{\alpha \in \mathbb{R}}\{R(\alpha,\beta)\} = \beta + 1.
\]
In a similar way, the optimal value can be got in each case.
Summarizing them, we obtain the proximity operator of $g$:
\[
 \{\prox{g}{z}\}_i = \left\{
 \begin{array}{ll}
 z_i + 1 &\mathrm{if}\mbox{ }z_i<-2, \\
 -1 &\mathrm{if}\mbox{ } {-2} \leq z_i < -1, \\
 z_i &\mathrm{if}\mbox{ } {-1} \leq z_i < 1, \\
 1 &\mathrm{if}\mbox{ } 1 \leq z_i < 2, \\
 z_i-1 &\mathrm{if}\mbox{ } 2 \leq z_i. \\
 \end{array}\right.
\]
\end{IEEEproof}
With the proximity operator, the following algorithm is FISTA to solve (\ref{soav2}):
\begin{algorithm}[FISTA~\cite{Beck2009}]
Fix $\tilde{z}^{(1)} \in \mathbb{R}^N,t_1=1$ and $L \in \mathbb{R}$ which is greater than or equal to a Lipshitz constant of $\nabla f$.
For $k \geq 1$,
\[
 \left\{
 \begin{array}{rcl} \vspace{.2em}
 z^{(k)} & = & \displaystyle{\prox{\frac{1}{L}g}{\tilde{z}^{(k)}-\frac{1}{L} \nabla f(z)},} \\ \vspace{.2em}
 t_{k+1} & = & \displaystyle{\frac{1+\sqrt{1+4t_{k}^2}}{2},} \\
 \tilde{z}^{(k)} & = & \displaystyle{z^{(k)} + \left( \frac{t_k-1}{t_{k+1}} \right) \left(z^{(k)} - z^{(k-1)}\right).} \\
 \end{array}
 \right.
\]
\end{algorithm}
It is known that $z^{(k)}$ converges to a solution of the optimization problem.
Note that here $\nabla f(z)$ can be calculated as $2\lambda H^{\top}(Hz-y)$.
Finally, with the obtained optimal solution $z^{\ast}$ we estimate the original signal by
\[
 x_{\mathrm{est}} = \mathrm{sign}(z^{\ast}),
\]
where $\mathrm{sign}(\cdot)$ is the signum function.

\section{Simulation results}
\label{SR}

In this section, we give some simulation results to demonstrate the performance of the proposed method.

In the simulations we employ QPSK for digital modulation.
We model the components of a modulation matrix $\tilde{H} \in \mathbb{C}^{M \times N}$ as
i.i.d. Gaussian variables with zero mean and variance $1/M$.
It is assumed that the elements of the symbol vector $x$ are independently and uniformly distributed with equal probability.
We set $\lambda,L,$ and $\tilde{z}^{(1)}$ to be $0.01,0.1,$ and $1_N$ respectively.
The maximum number of iterations of FISTA is $100$.

\begin{figure}[t]
\centering
\includegraphics[width=0.98\linewidth]{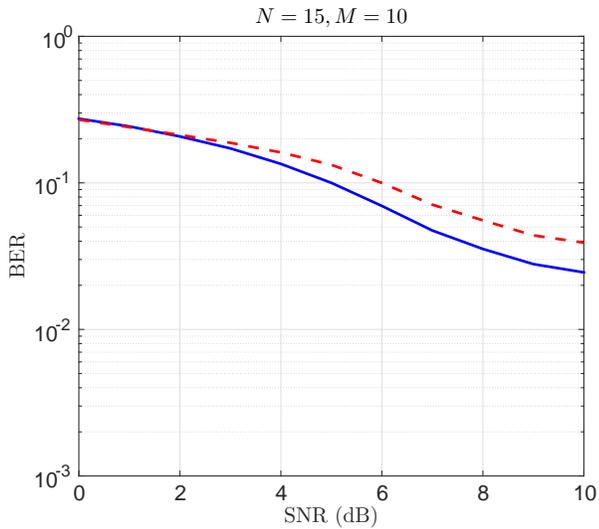}~
\caption{BER performance of the proposed detector (solid) and the $\ell^{\infty}$ minimization detector (broken) when $N=15,M=10$.}
\label{fig:berN15}
\end{figure}

\begin{figure}[t]
\centering
\includegraphics[width=0.98\linewidth]{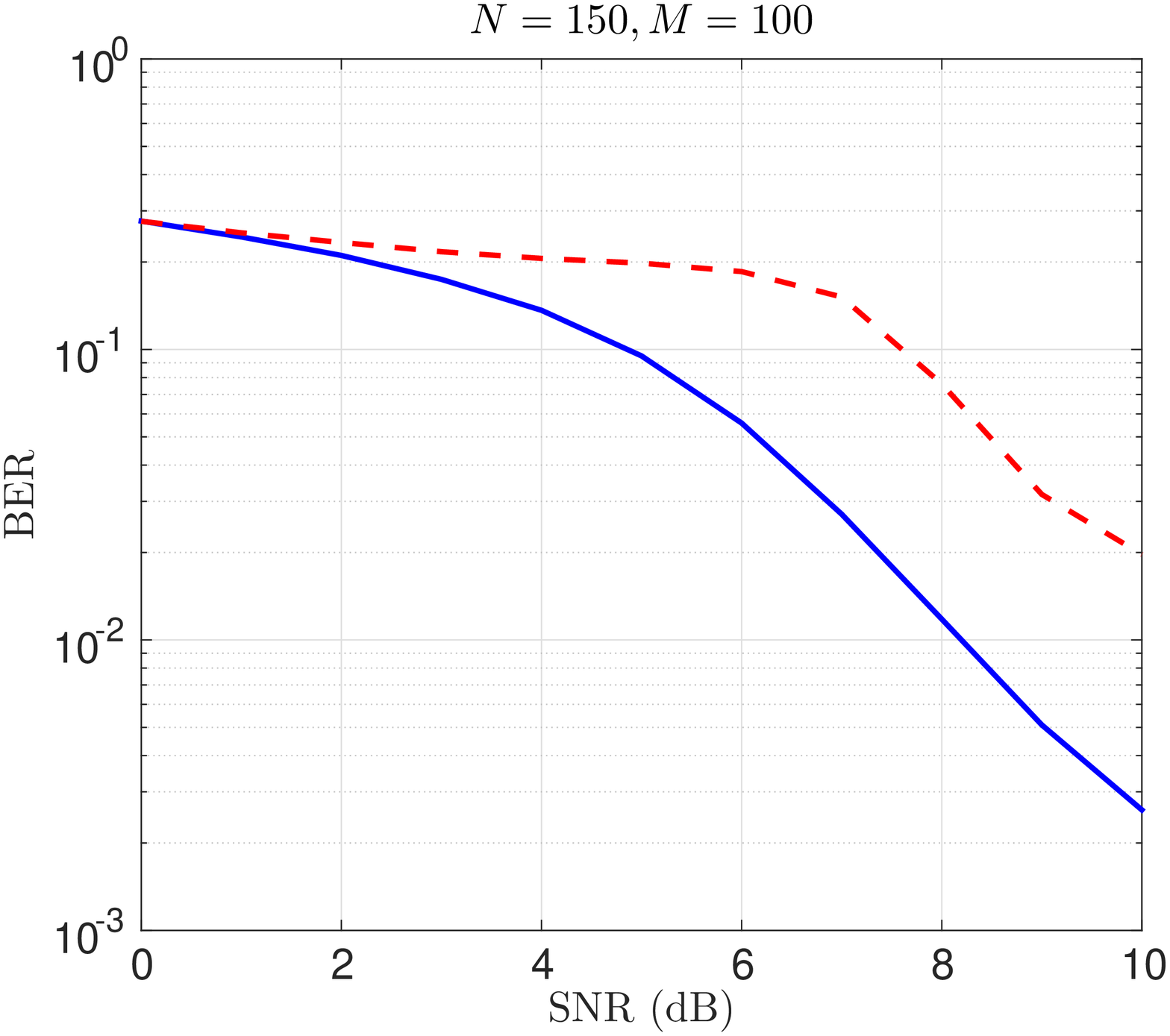}~
\caption{BER performance of the proposed detector (solid) and the $\ell^{\infty}$ minimization detector (broken) when $N=150,M=100$.}
\label{fig:berN150}
\end{figure}

Fig.~\ref{fig:berN15} and Fig.~\ref{fig:berN150} show the bit error rate (BER) performance against signal-to-noise ratio (SNR) when $N=15,M=10$ and $N=150,M=100$ respectively.
That is, the dimensions of the real valued matrix $H$ become $30 \times 20$ and $300  \times 200$.
The solid lines represent the performances of the proposed detector and the broken lines represent the performances of the $\ell^{\infty}$ minimization detector~\cite{Fang2014}.
BER performance is obtained by averaging BERs for $1000$ realizations of the modulation matrix for each SNR with the transmission of $900$ symbols for each realization.
From these figures we can see that, while the performance of the proposed method is slightly better than the $\ell^{\infty}$ minimization method with the small modulation matrix,
considerable performance gain can be achieved with the proposed scheme for the large $H$.

Next, we compare the proposed method using FISTA with the $\ell^{\infty}$ minimization method using the Newton's method in terms of computation time~\cite{Fang2014}.
Table~\ref{tab1} shows the average computation times to solve one optimization problem by the methods when $N=150,M=100$.
In the simulations, we use a computer with Intel Core i5-4590 CPU.
The results show that the computational complexity of the proposed scheme is lower than the conventional scheme, while achieving better BER performance.

\begin{table}[t]
\renewcommand{\arraystretch}{1.3}
\caption{Computation times comparison}
\label{tab1}
\centering
 \begin{tabular}{l|l} \hline
 Proposed Method & 0.002979 [sec] \\ \hline
 $\ell^{\infty}$ Minimization & 0.015251 [sec] \\ \hline
 \end{tabular}
\end{table}

\section{Conclusion}
\label{conc}

In this letter, we have proposed a symbol detection method for faster-than-Nyquist singling system by SOAV with FISTA.
We have derived the proximity operator in the SOAV optimization for binary symbol detection.
We have also shown simulation results to illustrate the effectiveness of the proposed method.
%\textcolor{magenta}{It is notable that the proposed method can be applied to multi-level baseband modulation systems.}

%\appendix[Derivation of $\prox{g}{z}$]
% or


%\appendices

%\section{}

\section*{Acknowledgment}
This research was supported in part by JSPS KAKENHI Grant Numbers
15H02668, 15K14006, 26120521, 15K06064, and 15H02252.

\ifCLASSOPTIONcaptionsoff
  \newpage
\fi

\bibliographystyle{IEEEtran}% bib style
\bibliography{sshrrefs}
%\bibliography{IEEEabrv,sshrrefs}% your bib database

\end{document}